\documentclass[final,3p,times]{elsarticle}
\usepackage{amsmath}
\usepackage{amsthm}
\usepackage{amsfonts}
\usepackage{lineno,hyperref}
\usepackage{mathrsfs}
\usepackage{graphicx}
\usepackage{adjustbox}
\usepackage{xcolor,colortbl}
\usepackage{float}
\usepackage{subfig}
\usepackage{siunitx}
\restylefloat{table}
\usepackage[T1]{fontenc}
\usepackage{array}
\usepackage{makecell}
\newcolumntype{x}[1]{>{\centering\arraybackslash}p{#1}}

\usepackage{tikz}
\newcommand\diag[4]{%
  \multicolumn{1}{p{#2}|}{\hskip-\tabcolsep
  $\vcenter{\begin{tikzpicture}[baseline=0,anchor=south west,inner sep=#1]
  \path[use as bounding box] (0,0) rectangle (#2+2\tabcolsep,\baselineskip);
  \node[minimum width={#2+2\tabcolsep},minimum height=\baselineskip+\extrarowheight] (box) {};
  \draw (box.north west) -- (box.south east);
  \node[anchor=south west] at (box.south west) {#3};
  \node[anchor=north east] at (box.north east) {#4};
 \end{tikzpicture}}$\hskip-\tabcolsep}}

\definecolor{Gray}{gray}{0.85}

\newtheorem{exm}{Example}
\modulolinenumbers[5]

\begin{document}

\title{ A Numerical Method for Pricing Discrete Double Barrier Option by Lagrange Interpolation on Jacobi Node }

\author[label1]{Amirhossein Sobhani\corref{cor1}}
\ead{a\_sobhani@aut.ac.ir, a\_sobhani@mathdep.iust.ac.ir}
\address[label1]{Department of Applied Mathematics, Amirkabir University of Technology, No. 424, Hafez Ave., Tehran 15914, Iran}
\cortext[cor1]{Corresponding author}
\author[label2]{Mariyan Milev}
\address[label2]{UFT-PLOVDIV, Department of Mathematics and Physics}
\ead{marianmilev2002@gmail.com}

\begin{frontmatter}

\begin{abstract}
In this paper, a rapid and high accurate numerical method for pricing discrete single and double barrier knock-out call options is presented. According to the well-known Black-Scholes framework, the price of option in each monitoring date could be calculate by  computing a recursive integral formula upon the heat equation solution. We have approximated these recursive solutions with the aim of Lagrange interpolation on Jacobi polynomials node. After that, an operational matrix, that makes our computation significantly fast, has been driven. The most important feature of this method is that its CPU time dose not increase when the number of monitoring dates increases. The numerical results confirm the accuracy and efficiency of the presented numerical algorithm. 
\end{abstract}

\end{frontmatter}

\section{Introduction}
Barrier options play a key role in financial markets where the most important problem is the so called option valuation problem, i.e. to compute a fair value for the option, i.e. the premium. The Nobel Prize-winning Black-Scholes option valuation theory motivates using classical numerical methods for partial differential equations (PDE's) \cite{smith1985numerical}. In computational Finance numerous nonstandard numerical methods are proposed and successfully applied for pricing options \cite{duan2003pricing,fusai2006,fusai2007analysis,milev2010numerical,tagliani2013laplace,
gzyl2017discontinuous,sobhani2018numerical,dilloo2017high,milev2013efficient,ndogmo2007high,kabaivanov2017modelling}. Numerical methods are often preferred to closed-form solutions as it they could me more easily extended or adapted to satisfy all the financial requirements of the option contracts and continuously changing conditions imposed by financial institutions and over-the-counter market for controlling trading of derivatives.
\par
Kunitomo and Ikeda \cite{kunitomo1992pricing} obtained general pricing formulas for European double barrier options with curved barriers but like for a variety of path-dependent options and corporate securities most formulas are obtained for restricted cases as continuous monitoring or single barrier \cite{milev2010numerical}. The discrete monitoring is essential as the trading year is considered to consist of $250$ working days and a week of 5 days. Thus, taking for one year $T=1$, the application of barriers occurs with a time increment of $0.004$ daily and $0.02$ weekly.
\par
For discrete barrier options there are some analytical solutions. For example, Fusai reduces the problem of pricing one barrier option to a Wiener-Hopf integral equation \cite{fusai2006}. Several other different contracts with discrete time monitoring are characterized by updating the initial conditions, such as Parisian options and occupation time derivatives \cite{fusai2002practical}. We remark that although most real contracts specify fixed times for monitoring the asset, academic researchers have focused mainly on continuous time monitoring models as the analysis of fixed barriers could be treated mathematically using some techniques such as the reflection principle \cite{kwokmathematical}. For example, using the reflection principle in Brownian motions, Li expresses the solution in general as summation of an infinite number of normal distribution functions for standard double barrier options, and in many non-trivial cases the solution consists of finite terms \cite{li1999pricing}. Pelsser derives a formula for continuous double barrier knock-out and knock-in options by inverting analytically the Laplace transform by a contour integration, \cite{pelsser2000pricing}. Broadie et. al. have found an explicit correction formula for discretely monitored option with one barrier \cite{broadie1997continuity}. However, these three well-known methods \cite{tagliani2013laplace,milev2013efficient,ndogmo2007high} have not been still applied in the presence of two barriers, i.e. a discrete double barrier option.
\par
Although it could not be claimed that it is impossible to be found an exact or closed-form solution of the Black-Scholes equation \cite{barndorff1997processes} for the valuation of discrete double barrier knock-out call option, it is sure that there is a substantial differences in the option prices between continuous and discrete monitoring even for 1 000 000 monitoring dates. This could be trivially tested for a single barrier knock-in and knock-out option using formulas \cite{fusai2006}, \cite{kunitomo1992pricing}[6], or the correction formula \cite{broadie1997continuity}, for double barrier knock-out options with the numerical algorithm \cite{milev2010numerical} or with a high-order accurate finite difference scheme \cite{ndogmo2007high}. It is well-known in literature the relation when comparing the price of continuous and discretely monitored barrier options with the corresponding vanilla option with same parameters and absence of rebates. The discrete monitoring considerably complicates the analysis of barrier options \cite{broadie1997continuity} and their pricing often requires nonstandard method as those presented in \cite{duan2003pricing,milev2010numerical,gzyl2017discontinuous,ndogmo2007high}. Difficulties of pricing double barrier options emerge even in the case of continuous monitoring where some drawbacks of close-form formulas could be clearly observed. The analytical solutions of such options is usually expressed as infinite series of reflections and presented with Fourier series. For fixed barriers contracts the Fourier series solution gives the same answer when all the terms have been added up but the main drawback is that the rate of convergence of the sum to the solution can be quite different, depending on the time to expiry.
\par
Initially classical quantitative methods in Finance have been explored for pricing barrier options. This includes standard lattice techniques, i.e. the binomial and trinomial trees of Kamrad and Ritchken \cite{kamrad1991multinomial}, Boyle and Lau \cite{boyle1994bumping}, Kwok \cite{kwokmathematical}, Heyen and Kat \cite{heynen1997barrier}, Tian \cite{tian1999pricing}, Dai and Lyuu \cite{dai2010bino} used standard lattice techniques, the binomial and trinomial trees, for pricing barrier options. Ahn et al. \cite{ahn1999pricing} introduce the adaptive mesh model (AMM) that increases the efficiency of trinomial lattices. The Monte Carlo simulation methods were implemented in \cite{andersen1996exact,beaglehole1997going,baldi1999pricing, bertoldi2003monte,kuan2003pricing,jeong2017hybrid}.  Also numerical algorithms based on quadrature methods have been proposed in \cite{andricopoulos2003universal,milev2010numerical}.
\par
Recently a great variety of more sophisticated semi-analytical methods for pricing barrier options have been developed which are based on integral transforms \cite{fusai2006,broadie2005double,fang2009pricing}, or on the transition probability density function of the process used to describe the underlying asset price \cite{andricopoulos2003universal,milev2010numerical,golbabai2014highly,broadie1997continuity,
dorfleitner2008pricing,fusai2007analysis,skaug2007fast,sullivan2000pricing}. Farnoosh et al. \cite{farnoosh2015,farnoosh2015a} have proposed a numerical algorithms for pricing discrete single and double barrier options with time-dependent parameters, while in \cite{amir1} a projection methods have been explored. These techniques are very high performing for pricing discretely monitored single and double barrier options and our computational results are in very good agreement with them. The main objective of this paper is present a new efficient computational method for valuation of discrete barrier options based on a Lagrange interpolation on Jacobi nodes that have not only a simpler computer implementation but also differ with minimum memory requirements and extreme short computational times.
\par
This article is organized as follows. In Section 2 we formulate the mathematical model for valuation of barrier options under the classical Black-Scholes framework. In Section 3 we briefly list definitions for Jacobi Polynomials. In section 4 we propose a new efficient numerical methods where an orthogonal Lagrange interpolation is utilized and a suitable operational matrix form has been obtained for pricing discrete double barrier options. One of the main advantages of this algorithm is that it do not depend on the number of monitoring dates. In the next Section 5 we observe numerical errors of order $10^{-4}$ and $10^{-6}$ in maximum norm for different computational experiments according to the number of node points. The obtained results are in good agreement with other benchmark values in literature and this confirms the efficiency and accuracy of the presented numerical algorithm.
\section{The Pricing Model}
We assume that the stock price process $S_t$ follows the Geometric Brownian motion:
	\[ \frac{d{S_t}}{S_t}  = r dt + \sigma d{B_t} \]
where \({S_0}\), \({r}\) and \(\sigma\) are  initial stock price, risk-free rate and volatility respectively. 
 We consider the problem of pricing knock-out discrete double barrier call option, i.e. a call option that becomes worthless if the stock price touches either lower or upper barrier at the predetermined monitoring dates: \[0 = {t_0} < {t_1} <  \cdots  < {t_M} = T.\]
  We assume that that monitoring dates are  equally spaced, i.e; \({t_m} = m\tau \) where $\tau =\frac{T}{M}$. If the barriers are not touched in monitoring dates, the pay off at maturity time is $\max(S_T-E,0)$, where $E$ is exercise price. The price of option is defined discounted expectation of pay off at the maturity time. Based on the Black-Scholes framework, the option price \(\mathcal{P}\left( {S,t,m - 1} \right)\) as a function of stock price at time \(t \in \left( {{t_{m - 1}},{t_m}} \right)\), satisfies in the following partial differential equations 
  \begin{equation} \label{pde:1}
 - \frac{{\partial \mathcal{P}}}{{\partial t}} + r S\frac{{\partial \mathcal{P}}}{{\partial S}} + \frac{1}{2}{\sigma ^2}{S^2}\frac{{{\partial ^2}\mathcal{P}}}{{\partial {S^2}}} - r\mathcal{P} = 0,
\end{equation} 
subject to the initial conditions:
  \[ \mathcal{P}\left( {S,{t_0},0} \right) = \left( {S - E} \right){\textbf{1}_{\left(\max (E,L) \le S \le U\right)}}\,\]
	\[ \mathcal{P}\left( {S,{t_m},0} \right) = \mathcal{P}\left( {S,{t_m},m - 1} \right){\textbf{1}_{\left(L \le S \le U\right)}};\,\,m = 1,2,...,M - 1~,\]
where \(\mathcal{P}\left( {S,{t_m},m - 1} \right): = \mathop {\lim }\limits_{t \to {t_m}} \mathcal{P}\left( {S,t,m - 1} \right)\).

By denoting \({E^*} = \ln \left( {\frac{E}{L}} \right);\,\mu  = {r} - \frac{{{\sigma ^2}}}{2};\,\theta  = \ln \left( {\frac{U}{L}} \right)\) and \( \delta  = \max \left\{ {{E^*},0} \right\}\), we define \({\mathrm{g}_m}\left( z \right)\) as following recursive formula:	
\begin{equation}
\label{f1b}
{\mathrm{g}_1}(z) = \int_0^\theta  {{k}(z - \xi ,\tau){\mathrm{g}_0}\left( \xi  \right)d\xi } 
\end{equation}
\begin{equation}
\label{fnb}
{\mathrm{g}_m}(z) = \int_0^\theta  {{k}(z - \xi ,\tau){\mathrm{g}_{m - 1}}\left( \xi  \right)d\xi } ;\,m = 2,3,...,M
\end{equation}	
where
\begin{equation}
\label{f0b}
\mathrm{g}_0\left( z  \right) = L{e^{ - \alpha z}}\left( {{e^z} - {e^{{E^*}}}} \right){\textbf{1}_{\left( \delta  \le z  \le \theta \right)}},
\end{equation}
\begin{equation}
\label{k1}
k(z,t) = \frac{1}{{\sqrt {4\pi {c^2}t} }}{e^{ - \frac{{{z^2}}}{{4{c^2}t}}}}.
\end{equation}
It could be shown that the price of the knock-out discrete double barrier option can be obtain as follows ( see \cite{amir1} ):
	\begin{equation}
	\mathcal{P}\left( {S_0,t_M,M - 1}\right) \simeq {e^{\alpha z_0 + \beta t}}{ \mathrm{g}_{M} \left( z_0/\theta  \right) }
	\end{equation}
where $z_0=\log\left( \frac{S_0}{L}\right)$.
\section{Jacobi Polynomials}
Let $w^{(\alpha,\beta)}(x)=(1-x)^\alpha (1+x)^\beta,~ \alpha,\beta > -1,$ and $L^2_{w^{(\alpha,\beta)}}(-1,1)$ be Hilbert space with the following inner product and norm:
\begin{equation}
 <f,g>_{w^{(\alpha,\beta)}}=\int_{-1}^1{f(x)g(x)w(x)dx},
 \end{equation} 
 \begin{equation}
\Vert f \Vert_{w^{(\alpha,\beta)}} =\sqrt{<f,f>_{w^{(\alpha,\beta)}}}.
 \end{equation}
 
 The Jacobi polynomials, $J_i^{(\alpha,\beta)}(x)$ are orthogonal polynomials in $L^2_{w^{(\alpha,\beta)}}(-1,1)$, i.e;
 \begin{equation}
 \int_{-1}^1{J_i^{(\alpha,\beta)}(x)J_j^{(\alpha,\beta)}(x)w(x)dx}=\lambda_{i}\delta_{ij},
 \end{equation}
 where $\lambda_{i}= \Vert {J}_i^{(\alpha,\beta)} \Vert^2$. These polynomials, that set an orthogonal basis in $L^2_{w^{(\alpha,\beta)}}(-1,1)$, satisfy in following three-term recurrence relation:

\begin{eqnarray}
J_0^{(\alpha,\beta)}(x)=1,~J_1^{(\alpha,\beta)}(x)=\frac{1}{2}(\alpha+\beta+2)x+\frac{1}{2}(\alpha-\beta)\\
J_{i+1}^{(\alpha,\beta)}(x)=\left( a_i^{(\alpha,\beta)}x-b_i^{(\alpha,\beta)} J_{i}^{(\alpha,\beta)}(x) \right)-c_i^{(\alpha,\beta)} J_{i-1}^{(\alpha,\beta)}(x)
\end{eqnarray}
where:
\begin{eqnarray}
a_i^{(\alpha,\beta)}=\frac{(2i+\alpha+\beta+1)(2i+\alpha+\beta+2)}{2(i+1)(n+\alpha+\beta+1)}\\
b_i^{(\alpha,\beta)}=\frac{(\beta^2-\alpha^2)(2n+\alpha+\beta+1)}{2(i+1)(n+\alpha+\beta+1)(2n+\alpha+\beta)}\\
c_i^{(\alpha,\beta)}=\frac{(n+\alpha)(n+\beta)(2n+\alpha+\beta+2)}{(i+1)(n+\alpha+\beta+1)(2n+\alpha+\beta)}.
\end{eqnarray}

\section{Pricing by orthogonal Lagrange interpolation}
In this section we consider ${{\Pi }_{n}}$ as space of all polynomials with degree less or equal to $n$, set points $\{{{x}_{i}^{\alpha,\beta}}\}_{i=0}^{n}$ as roots of $(n+1)$-th Jacobi  polynomial ${J}_{n+1}^{(\alpha,\beta)}$ that are shifted to $[0,\theta ]$ and ${\mathcal{I}_{n}^{\alpha,\beta}}:C[0,\theta ]\to {{\Pi }_{n}}$  as orthogonal polynomial interpolation projection operator, that is defined as follows:
	\begin{equation}
	 	\mathcal{I}_{n}^{\alpha,\beta}\left( f \right)=\sum\limits_{i=0}^{n}{f({{x}_{i}^{\alpha,\beta}})}{\mathcal{L}_{i}}(x)
	\end{equation} 	
where ${\mathcal{L}_{i}}(x)$ is the $i$-th Lagrange polynomial basis function defined on $\{{{x}_{i}^{\alpha,\beta}}\}_{i=0}^{n}$:
\begin{equation}
{\mathcal{L}_{i}}(x)=\prod\limits_{j=0,j\ne i}^{n}{\frac{(x-{{x}^{\alpha,\beta}_{j}})}{({{x}^{\alpha,\beta}_{i}}-{{x}^{\alpha,\beta}_{j}})}}.
\end{equation}
	
Let operator $\mathcal{K}:L^{2}([0,\theta]) \to L^{2}([0,\theta ])$ is defined as follows:
\begin{equation}
\label{operator}
\mathcal{K}\left( \mathrm{g} \right)(z): = \int_0^\theta  {\kappa (z - \xi ,\tau )\mathrm{g}(\xi )} d\xi .
\end{equation}
where $\kappa$ is defined in \eqref{k1}. 
According to the definition of operator $\mathcal{K} $, equations \eqref{f1b} and \eqref{fnb} can be rewritten as below:
\begin{equation}
\mathrm{g}_1=\mathcal{K}\mathrm{g}_{0}
\end{equation}
\begin{equation}
\mathrm{g}_m=\mathcal{K}\mathrm{g}_{m-1}~~\,m = 2,3,...,M
\end{equation}
We denote
\begin{equation}
{{\tilde{\mathrm{g}}}_{1,n}} = {\mathcal{I}_{n}^{\alpha,\beta}}\mathcal{K}\left( {{\mathrm{g}_0}} \right)
\end{equation}
\begin{equation}
\label{recfmadm}
{{\tilde{\mathrm{g}}}_{m,n}} = {\mathcal{I}_{n}^{\alpha,\beta}}\mathcal{K}\left( {{\tilde{\mathrm{g}}}_{m - 1}} \right)= \left({\mathcal{I}_{n}^{\alpha,\beta}}\mathcal{K}\right)^{m}\left(\mathrm{g}_{0}\right)\,,\,m \ge 2.
\end{equation}
where ${\mathcal{I}_{n}^{\alpha,\beta}}\mathcal{K}$ is as follows:
\[({\mathcal{I}_{n}^{\alpha,\beta}}\mathcal{K})(\mathrm{g}) = {\mathcal{I}_{n}^{\alpha,\beta}}\left( {\mathcal{K}(\mathrm{g})} \right).\]
Since, \({\tilde{\mathrm{g}}}_{m,n} \in {{\Pi }_{n}}\) for \( m \geq 1\), we can write 
	\[{\tilde{\mathrm{g}}}_{m,n} = \sum\limits_{i = 0}^{n} {{a_{mi}}{\mathcal{L}_i}(z)}  = {\Phi '_n}(x)G_{m},\] 	
where \({ G_m} = [{a_{m0}},{a_{m1}}, \cdots ,{a_{m{n}}}]'\) and \({ \Phi_n} = [{\mathcal{L}_{m}},{\mathcal{L}_{m}}, \cdots ,{\mathcal{L}_n}]'\).
From equation \eqref{recfmadm} we obtain
\begin{equation}
\label{19}
{\tilde{\mathrm{g}}}_{m,n} = {({\mathcal{I}_{n}^{\alpha,\beta}}\mathcal{K})^{m - 1}} \left( {\tilde{\mathrm{g}}}_{1,n} \right).
\end{equation}
Since \({\Pi }_{n}\) is a finite dimensional linear space, thus the linear operator \({\mathcal{I}_{n}^{\alpha,\beta}}\mathcal{K}\) on \({\Pi }_{n}\) could be considered as a \({n} \times {n}\) matrix \(K\). Consequently equation \eqref{19} can be written as following matrix operator form
\begin{equation}
\label{fmatrix}
{\tilde{\mathrm{g}}}_{m,n} = {\Phi '_n}{K^{m - 1}}{ \mathrm{G}_1}.
\end{equation}

For evaluation of the option price by \eqref{fmatrix}, it is enough to calculate the matrix operator \(K\) and the vector \({ G_1}\). It is easy to check (see \cite{amir1}) that:
\[{G_1} = [{a_{11}},{a_{12}}, \cdots ,{a_{1n}}]' \]
\[K = {\left( {{k_{ij}}} \right)_{{n} \times {n}}}\]
where
\[{a_{1i}} = \int_\delta^\theta {\kappa ({{{x}_{i}^{\alpha,\beta}}  - \xi ,\tau ){\mathrm{g}_0}(\xi )d\xi  } \,\,} ,\,\,0 \le i \le n.\]
\[{k_{ij}} = \int_0^\theta  \kappa ({{{x}_{i}^{\alpha,\beta}}  - \xi ,\tau ) {{\mathcal{L}_{j-1}}(\xi ) d\xi } \,\,} .\]
Therefore, the price of the knock-out discrete double barrier option can be estimated as follows:
	\begin{equation}
	\label{fNmatrix}
	\mathcal{P}\left( {S_0,t_M,M - 1}\right) \simeq {e^{\alpha z_0 + \beta t}}{{\tilde{\mathrm{g}}}_{M,n} \left( z_0 \right) }
	\end{equation}
where $z_0=\log\left( \frac{S_0}{L}\right)$ and ${\tilde{\mathrm{g}}}_{M,n}$ from \eqref{fmatrix}.
The matrix form of relation \eqref{fmatrix} implies that the computational time of presented algorithm be nearly fixed when monitoring dates increase. Actually, the complexity of our algorithm is $\mathcal{O}{(n^2)}$ that dose not depend on number of monitoring dates.
\section{Numerical Result}
In the current section, the presented method in previous section for pricing knock-out call discrete double barrier option is compared with some other methods. The numerical results are obtained from the relation \eqref{fNmatrix} with $n$ basis functions. The Source code has been written in \textsc{Matlab} 2015 on a 3.2 GHz Intel Core i5 PC with 8 GB RAM.  
 \begin{exm}
 \label{example1}
 
 In the first example, the pricing of knock-out call discrete double barrier option is considered with the following parameters: $r=0.05$, $\sigma=0.25$, $T=0.5$, $S_0=100$, $E=100$, $U=120$ and $L=80~,90~,95~,99~,99.5$. In table \eqref{tab1}, numerical results of presented method with Milev numerical algorithm \citep{milev2010numerical}, Crank-Nicholson \cite{wade2007smoothing}, trinomial, adaptive mesh model (AMM) and quadrature method QUAD-K200 as benchmark \cite{shea2005numerical} are compared for various number of monitoring dates. In addition, it can be seen that CPU time of presented method is fixed against increases of monitoring dates. 
\end{exm}

\begin{table}[t]

\centering
\caption{The maximum norm error for $n=25$ of example\eqref{example1} with $L=95$ and $M=125$.}
\label{my-label}
\begin{tabular}{|c|c|c|c|c|c|}
\hline
 \diag{.1em}{1cm}{$\alpha$}{$\beta$} & $-0.8$ & $-0.5$  & 0 &  0.5 & 0.8\\ \hline
$-0.8$& $8.6074e-06$ & $8.1718e-06$ & $2.2103e-05$ & $1.8770e-05$ & $9.5120e-06$ \\ \hline
$-0.5$& $8.7606e-06$& $7.8929e-06$ & $ 1.2852e-05$ & $3.8600e-05$ & $4.6071e-05$ \\ \hline
0 &  $  2.7040e-05$& $2.5788e-05$ & $2.2103e-05$ & $5.3726e-05$ & $9.3608e-05$\\ \hline
0.5 &  $9.1438e-05$& $9.5461e-05$ & $9.2619e-05$ & $8.2564e-05$ & $1.0667e-04$ \\ \hline
0.8 &  $ 1.4675e-04$& $1.6600e-04$ & $ 1.7498e-04$ & $1.6536e-04$ & $1.5456e-04$\\ \hline
\end{tabular}
\end{table}

\begin{table}[t]
\caption{Double barrier option pricing of Example \eqref{example1}: $T = 0.5$, $r= 0.05$, $\sigma = 0.25$, $S_0 = 100$, $E = 100$.}
\label{tab1}
\footnotesize
\begin{center}

\begin{tabular} {c c c c c c c c }
\hline 
\multicolumn{1}{c}{M}  & \multicolumn{1}{c}l{L} &\begin{tabular}[c]{@{}c@{}}PM $(\alpha=-0.5,\beta=-0.5)$\\(n=25)\end{tabular} & \begin{tabular}[c]{@{}c@{}}Milev\\ ($200$)\end{tabular}& \begin{tabular}[c]{@{}c@{}}Milev\\ ($400$)\end{tabular} &\multicolumn{1}{c}{Trinomial}&\multicolumn{1}{c}{AMM-8}& \multicolumn{1}{c}{Benchmark}   \\ 
\hline 
& \multicolumn{5}{c}{} \\  
  & 80   &2.4499 &-	     &-       &2.4439 & 2.4499&2.4499 \\ 
  & 90   &2.2028 &-	     &-      &2.2717& 2.2027 &2.2028 \\ 
5 & 95   &1.6831 &1.6831 &1.6831 &1.6926&1.6830 &1.6831 \\ 
  & 99   &1.0811 &1.0811 &1.0811 &0.3153&1.0811 &1.0811 \\ 
  & 99.9 &0.9432 &0.9432 &0.9432 &-&0.9433 &0.9432 \\ 
\rowcolor[HTML]{C0C0C0} 
 CPU& \- & 0.035 s &1 s & 5 s &\multicolumn{3}{c}{}  \\ 
   & 80   &1.9420 &-      &-     &1.9490&1.9419 &1.9420 \\ 
   & 90   &1.5354 &-      &-     &1.5630&1.5353 &1.5354 \\ 
25 & 95   &0.8668 &0.8668 &0.8668 &0.8823&0.8668&0.8668 \\ 
   & 99   &0.2931 &0.2931 &0.2931 &0.3153&0.2932&0.2931 \\ 
   & 99.9 &0.2023 &0.2023 &0.2023 &-&0.2024&0.2023 \\ 
\rowcolor[HTML]{C0C0C0} 
 CPU& \- & 0.035 s& 8 s & 30 s &\multicolumn{3}{c}{}  \\ 
 & 80 & 1.6808 &- &- &1.7477&1.6807& 1.6808  \\ 
 & 90 & 1.2029 &- &- &1.2370&1.2028& 1.2029  \\ 
125& 95 & 0.5532 &0.5528 &0.5531  &0.5699&0.5531 & 0.5532  \\ 
& 99 & 0.1042 &0.1042 &0.1042  &0.1201&0.1043& 0.1042  \\ 
 & 99.9 & 0.0513 &0.0513 &0.0513&-&0.0513 & 0.0513  \\ 
\rowcolor[HTML]{C0C0C0} 
CPU& \- & 0.035 s& 35 s & 150 s &\multicolumn{3}{c}{}  \\ 

\end{tabular} 
\end{center}
\end{table}

\begin{figure}[H]
\subfloat[$M=125$]{
\includegraphics[width=.5\linewidth]{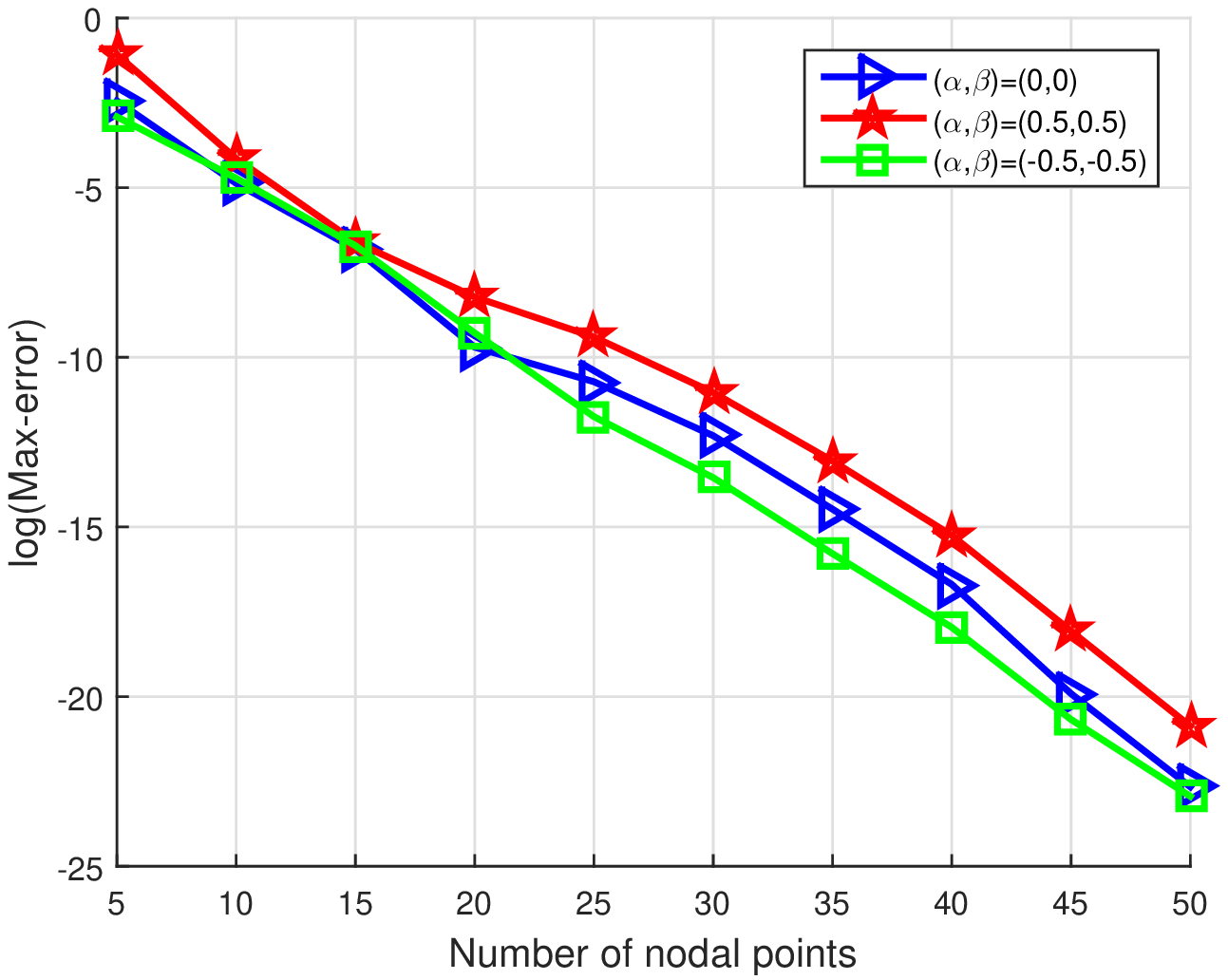}
}\hfill
\subfloat[$M=250$]{
\includegraphics[width=.5\linewidth]{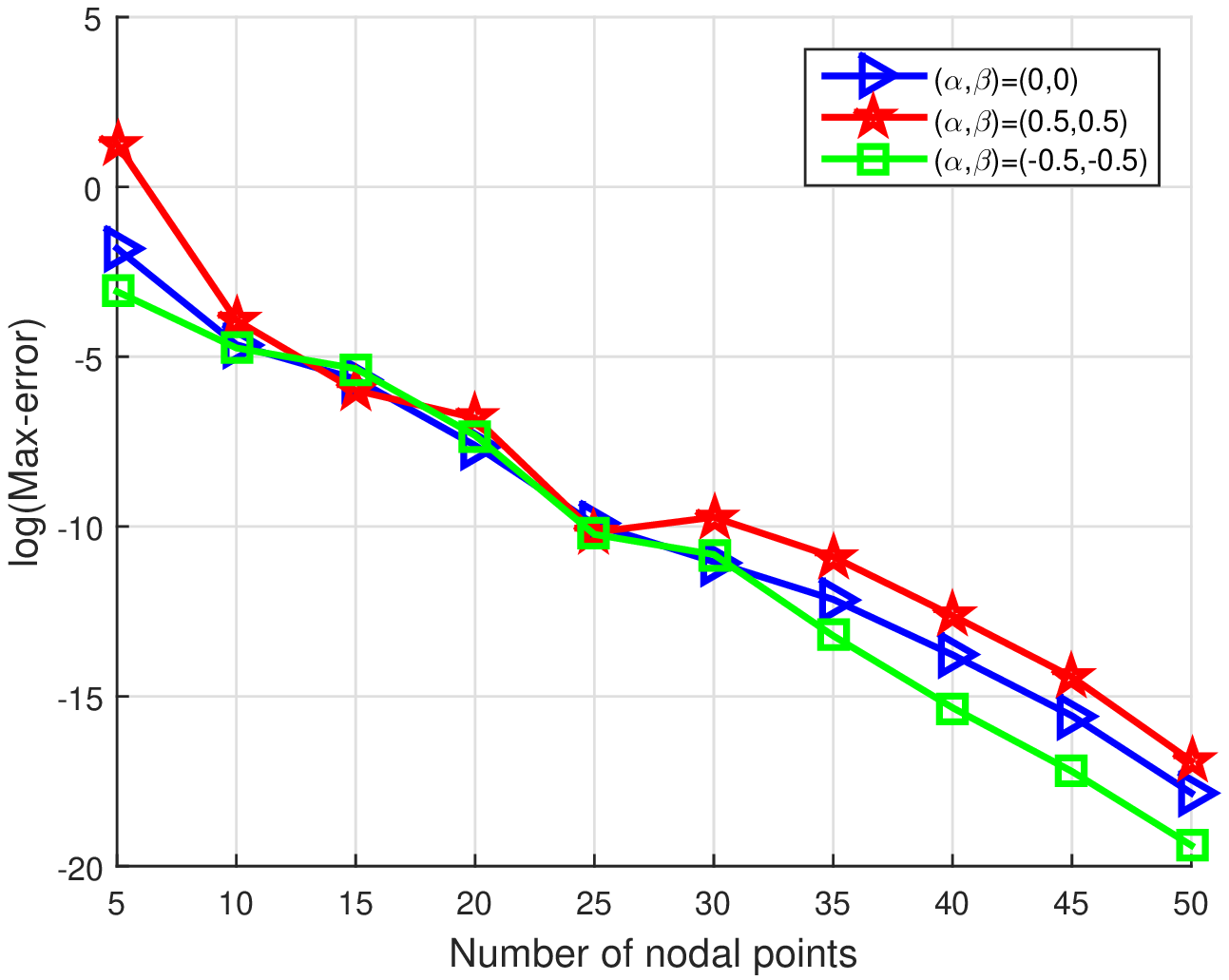}
}
\caption{$Max-error$ for example \eqref{example1} with $L=80$.}
\label{fig:1}
\end{figure}

\begin{figure}[htbp]
\subfloat[ Error]{
\includegraphics[width=.5\linewidth]{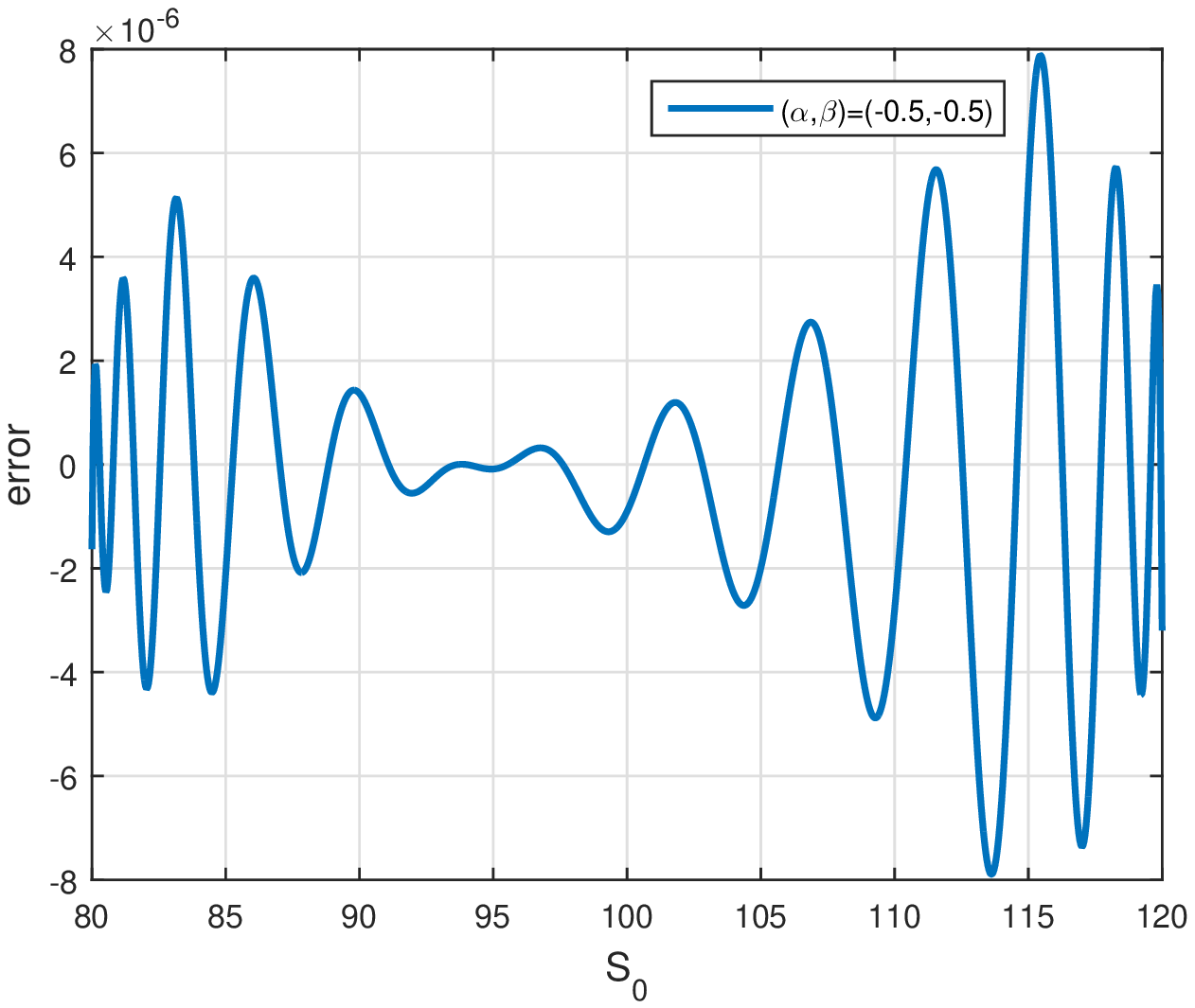}
}\hfill
\subfloat[ Estimated Price]{
\includegraphics[width=.5\linewidth]{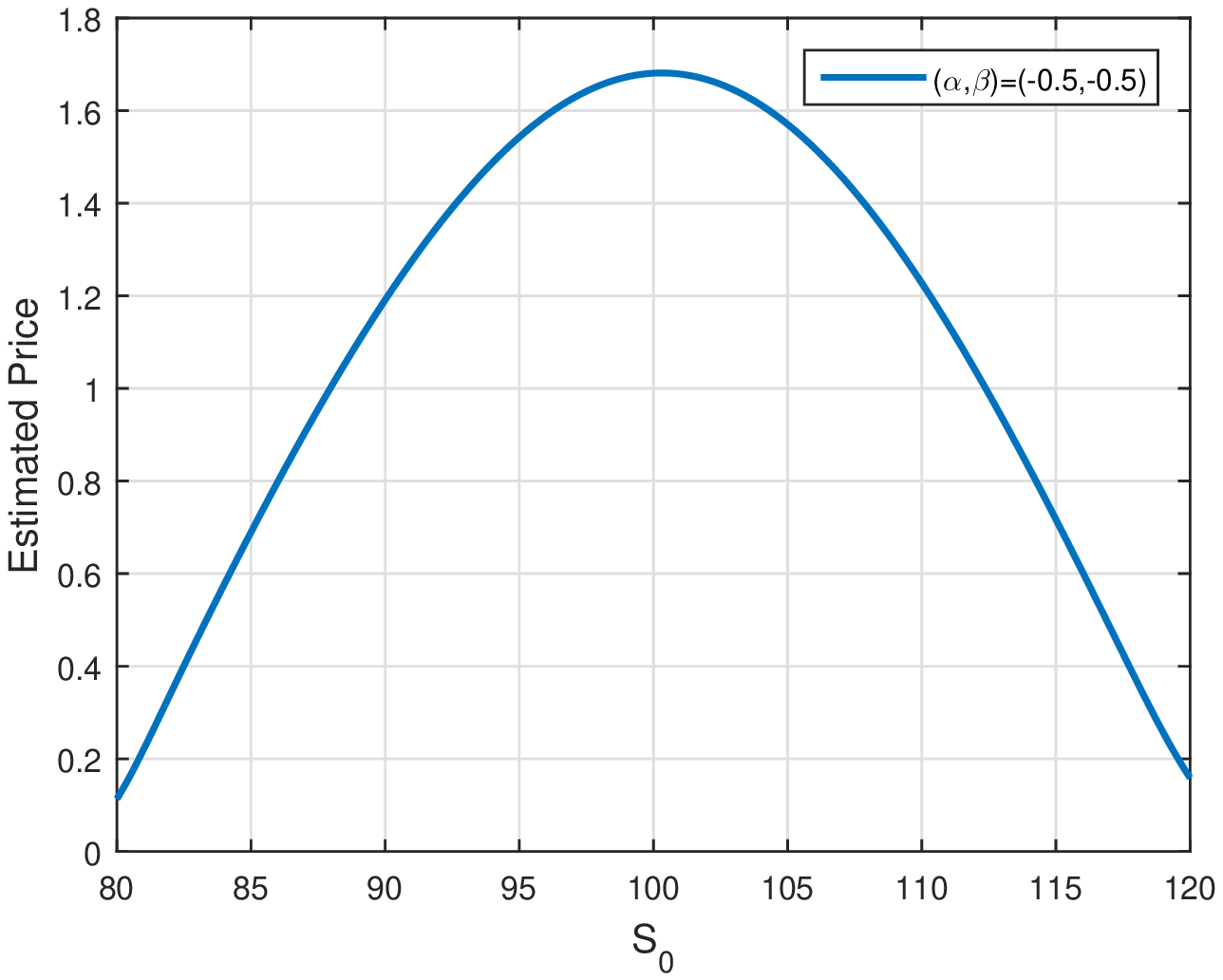}
}
\caption{The error and estimated Price in example\eqref{example1} with $L=80$ and $M=125$.}
\label{fig:2}
\end{figure}

\begin{exm}
\label{example2}
In this example, the parameters of knock-out call discrete double barrier option is considered as $r=0.05$, $\sigma=0.25$, $T=0.5$, $E=100$, $U=110$ and $L=95$. In table \eqref{exam2} the option price for different spot prices are evaluated and compared with Milev numerical algorithm \citep{milev2010numerical}, Crank-Nicholson \cite{wade2007smoothing} and the Monte Carlo (MC) method with $10^7$ paths \cite{brandimarte2003numerical}. 
\end{exm} 

\begin{table}[H]

\caption{Double barrier option pricing of Example \eqref{example2}: $T = 0.5$, $M=5$, $r= 0.05$, $\sigma = 0.25$, $E = 100$, $U=110$ and $L=95$.}
\label{exam2}
\resizebox{\columnwidth}{!}{%
\begin{tabular}{cccccc}
\hline
$S_0$    & \begin{tabular}[c]{@{}c@{}}PM $(\alpha=-0.5,\beta=-0.5,)$\\ $(n=25)$\end{tabular} & \begin{tabular}[c]{@{}c@{}} Crank-Nicolson \\ (1000) \end{tabular} & \begin{tabular}[c]{@{}c@{}}Milev\\ (400)\end{tabular} & \begin{tabular}[c]{@{}c@{}}Milev\\ (1000) \end{tabular}& \begin{tabular}[c]{@{}c@{}} MC (st.error) \\ with $10^7$ paths\end{tabular} \\ \hline
95       & 0.174498 & 0.1656 & 0.174503 & 0.174498& -\\
95.0001  & 0.174499 & $\simeq$ 0.1656 & 0.174501 & 0.174499 &0.17486 (0.00064)\\
95.5     & 0.182428 & 0.1732   & 0.182429& 0.182428 &0.18291 (0.00066)      \\
99.5     & 0.229349 & 0.2181 & 0.229356 & 0.229349 & 0.22923 (0.00073)\\
100      & 0.232508 & 0.2212  & 0.232514  & 0.232508& 0.23263 (0.00036) \\
100.5    & 0.234972 &0.2236   & 0.234978 & 0.234972 & 0.23410 (0.00073)    \\
109.5    & 0.174462 & 0.1658 & 0.174463 & 0.174462& 0.17426 (0.00063)\\
109.9999 & 0.167394 & $\simeq$ 0.1591 & 0.167399  & 0.167394 & 0.16732 (0.00062)   \\
110      & 0.167393 &0.1591 & 0.167398  & 0.167393 & -\\
\rowcolor[HTML]{C0C0C0} 
CPU      & 0.035 s & Minutes  & 1 s & 39 s&
\end{tabular}
}
\end{table}

\begin{exm}
\label{example3}
Due to the fact that the probability of crossing upper barrier during option's life when $U\geq2E$ is too small, the price of discrete single down-and-out call option can be estimated by double ones by setting upper barrier greater than $2E$ (for more details see\citep{milev2010numerical}). Now, we consider a discrete single down-and-out call option with the following parameters: $r=0.1$,  $\sigma=0.2$, $T=0.5$, $S_0=100$, $E=100$ and $L=95~,99.5~,99.9~$. The price is estimated by double ones with $U=2.5 E$. The numerical results are shown in table \eqref{exam3} and compared with Fusai's analytical formula \cite{fusai2006}, the Markov chain method (MCh)\cite{duan2003pricing} and the Monte Carlo method (MC) with $10^8$ paths \cite{bertoldi2003monte} that shows the validity of presented method in this case.
\end{exm}

\begin{table}[H]
\centering
\caption{Single barrier option pricing of Example \eqref{example3}: $T = 0.5$, $r= 0.1$, $\sigma = 0.2$, $S_0 = 100$, $E = 100$, $U=250$.}
\label{exam3}
\begin{tabular}{ccccccc}
\hline
  &  & \multicolumn{2}{c}{PM  $(\alpha=-0.5,\beta=-0.5)$} &  & &   \\ \cline{3-4}
L & M & n=25          & n=50        & (IR17) &    MCH                  & MC (st.error) \\ \hline
95   & 25  & 6.63104 & 6.63156 & 6.63156   &6.6307& 6.63204 (0.0009)                                                        \\
99.5 & 25  & 3.35644 & 3.35558 & 3.35558 & 3.3552& 3.35584 (0.00068)                                                           \\
99.9 & 25  & 3.00897 & 3.00887 & 3.00887 &3.0095&3.00918 (0.00064)                                                           \\
95   & 125 & 6.16940& 6.16863 & 6.16864 & 6.1678& 6.16879 (0.00088)                                                           \\
99.5 & 125 & 1.95811 & 1.96130 & 1.96130& 1.9617&1.96142 (0.00053) \\
99.9 & 125 & 1.50991 & 1.51020 & 1.51068 & 1.5138 & 1.5105 (0.00046) 
                                                          \\
 \rowcolor[HTML]{C0C0C0} 
CPU  &    &  0.038 s & 0.051 s& &                                                                                                                                                    
&                                                                  
\end{tabular}
\end{table}
\begin{exm}
\label{exm44}
In this example we estimate the price of continus monitoring call barrier down and out option, $\mathcal{P}c$, with discrete ones, $\mathcal{P}d_m$,  using the following formula\cite{broadie1997continuity}:
\[\mathcal{P}c(L)=\mathcal{P}d_m\left( L~e^{\beta \sigma \sqrt{\Delta t}}\right),\]
 where $\beta=\zeta(1/2)/\sqrt{2\pi}\simeq 0.5826$ with $\zeta$ the Riemann zeta function. The parameters of this problem is considered as $r=0.1$, $\sigma=0.3$, $T=0.2$, $E=100$, $S=100$.
 In table \eqref{ex4} the option price for different Lower barriers are evaluated and compared with continuous monitoring price that is obtained in \cite{broadie1997continuity}. As we can see, this estimations is accurate except when the barrier is close to the spot price.

\end{exm}

\begin{table}[H]
\centering
\caption{Single barrier option pricing with continuous monitoring of Example \eqref{exm44}: $T = 0.2$, $r= 0.1$, $\sigma = 0.3$, $S_0 = 100$, $E = 100$, $U=250$.}
\label{ex4}
\begin{tabular}{cccccc}
\hline
    &                    & \multicolumn{2}{c}{PM$(\alpha=-0.5,\beta=-0.5,M=50)$} & \multicolumn{2}{c}{PM$(\alpha=-0.5,\beta=-0.5,M=125)$} \\ \hline
L   & Countinous Barrier & n=25                     & n=50                     & n=25                      & n=50                     \\ \hline
85  & 6.308              & 6.307                    & 6.308                    & 6.306                     & 6.308                    \\
88  & 6.185              & 6.185                    & 6.185                    & 6.182                     & 6.185                    \\
91  & 5.808              & 5.808                    & 5.808                    & 5.809                     & 5.808                    \\
93  & 5.277              & 5.277                    & 5.277                    & 5.277                     & 5.277                    \\
95  & 4.398              & 4.396                    & 4.397                    & 4.398                     & 4.397                    \\
97  & 3.060              & 3.067                    & 3.067                    & 3.059                     & 3.059                    \\
99  & 1.171              & 1.479                    & 1.477                    & 1.265                     & 1.267                    \\
\rowcolor[HTML]{9B9B9B} 
CPU &                    & 0.038 s                  & 0.051 s                  & 0.038 s                   & 0.051 s                 
\end{tabular}
\end{table}
\section{Conclusion and remarks}
In this article, we used the Lagrange interpolation on Jacobi polynomial nodes for pricing discrete single and double barrier options. In section 4 we obtained a matrix relation \eqref{fmatrix} for solving this problem. Numerical results verify that computational time is fixed when the number of monitoring dates increase. 
 
\section*{References}

\bibliography{mybibfile}

\end{document}